\documentclass[12pt]{article}
\usepackage{graphicx}

\def\be{\begin{equation}} \def\ee{\end{equation}} \def\bea{\begin{eqnarray}}
\def\eea{\end{eqnarray}} \def\nnb{\nonumber}

\begin{document}

\hfill{July 6, 2026,\ \ \ {\tt n16Ov1.13}}

\begin{center}
\vskip 10mm 

\noindent
{\Large\bf  
	$S$ matrices of elastic $n$-$^{16}$O scattering at low energies
	in cluster effective field theory 
}
\vskip 10mm 

\noindent
{\large 
Shung-Ichi Ando\footnote{mailto:sando@sunmoon.ac.kr}, 
	
	\vskip 10mm
\noindent
{\it
Department of Display and Semiconductor Engineering
and Research Center for Nano-Bio Science, 
Sunmoon University,
Asan, Chungnam 31460,
Republic of Korea
	}%
}
\end{center}

\vskip 10mm

Elastic $n$-$^{16}$O scattering at low energies 
is studied in the framework of cluster effective field theory.
An evaluated data set of the total cross section 
of elastic $n$-$^{16}$O scattering at neutron energy, 
$0\le E_n\le 4$~MeV, is adopted 
from Evaluated Nuclear Data File (ENDF/B-VIII.0). 
We derive an expression for the $S$ matrices of the elastic scattering 
for seven spin-partial wave channels, 
$lj=s_{1/2}$, $p_{1/2}$, $p_{3/2}$, $d_{3/2}$, $d_{5/2}$, 
$f_{5/2}$, $f_{7/2}$, 
including one excited state and nineteen resonant states of $^{17}$O. 
Thirty-four parameters of the theory are fitted to the ENDF data, 
and we find that a plotted line, 
by using the fitted parameters, reproduces the ENDF data well. 
We discuss the uncertainties that may appear in the present approach
from the fitted parameters of the resonant states 
with widths, larger than $\Gamma = 90$~keV.
We also discuss the implications of the fitted values 
of energies and widths of the resonant states 
in the estimate of the astrophysical $S$ factor 
of $^{13}$C($\alpha$,$n$)$^{16}$O reaction at stellar energies.

\newpage 
\vskip 2mm \noindent
{\bf 1. Introduction}

Estimates of astrophysical $S$ factors of nuclear reactions
at low energies are essential for the study of stellar evolutions
and explosions and nucleosynthesis in stars~\cite{rr-88,i-15}.
$^{13}$C($\alpha$,$n$)$^{16}$O reaction plays an important role as 
a neutron source to create heavy elements through slow neutron captures,
the so-called $s$-process, in low-mass asymptotic giant branch (AGB) 
stars~\cite{dabps-epja26}. 
Recently, accurate measurements of $^{13}$C($\alpha$,$n$)$^{16}$O at 
low energies were reported from deep underground laboratories,
the LUNA collaboration~\cite{LUNA} and the JUNA collaboration~\cite{JUNA}.
In the previous work~\cite{sa-26}, 
we constructed an effective field theory (EFT) 
for the $^{13}$C($\alpha$,$n$)$^{16}$O reaction and extrapolated 
the $S$ factor to the stellar energies 
after fitting the parameters 
to the data reported by the LUNA and JUNA collaborations. 
We found that an uncertainty in the extrapolation appeared from a resonant
state of $^{17}$O around the $\alpha$-$^{13}$C breakup energy of $^{17}$O. 
The resonant states of $^{17}$O near the $\alpha$-$^{13}$C breakup threshold
can be investigated through elastic neutron-$^{16}$O scattering 
at low energies. 

Early study of the elastic $n$-$^{16}$O scattering
can be found in the 1950s,
e.g., in Refs~\cite{fc-pr58,wa-prc71,fjf-prc73}. 
This reaction is essential and selected as one of the highest-priority
isotopes for the CIELO collaboration~\cite{CIELO}, 
and the evaluated nuclear reaction data are compiled 
in the evaluated nuclear data files,
ENDF/B-VIII.0~\cite{betal-nds18}. 
Accurate knowledge of cross-section data and related uncertainties is also
crucial for the reactor analysis and design, and nuclear criticality safety
as well as for the processing and disposal of nuclear waste~\cite{skhn-ari18}. 
In this work, we adopt an evaluated data set of elastic $n$-$^{16}$O
scattering at low energies from ENDF/B-VIII.0 
as experimental data to fix parameters of the theory. 
One may refer to the compilation of resonant energies and 
widths of $^{17}$O in the TUNL nuclear data evaluation~\cite{twc-npa93}
as well. 

To construct an EFT for the study of a reaction 
is a popular theoretical method 
in hadron physics and few-body physics~\cite{w-physica79,hkvk-rmp20}.
When constructing an EFT, one needs to introduce a scale to 
separate the relevant degrees of freedom at low energy from irrelevant degrees of freedom at high energy. 
One constructs an effective Lagrangian with the fields of 
the relevant degrees of freedom, satisfying required symmetries,  
and perturbatively expand it in terms of the number of derivatives 
on the fields. 
Then, the reaction amplitudes can be 
calculated from the effective Lagrangian order by order, 
in powers of ${\cal Q}/\Lambda_H$, 
where ${\cal Q}$ is a typical momentum scale of a reaction in question,
and $\Lambda_H$ is the separation scale of the theory.
The irrelevant degrees of freedom at high energy are integrated out, 
and their effects are embedded 
in the coefficients of the terms of the effective Lagrangian.
Those coefficients, in principle, can be fixed from their mother theory,
but they are practically fitted to experimental data. 
The theory is also easy to 
include external electromagnetic and weak probes 
and to extend to the study of three-body systems. 
In the previous works, 
we constructed an EFT for $\alpha$-$^{12}$C system and
studied elastic $\alpha$-$^{12}$C
scattering~\cite{sa-epja16,sa-prc18,sa-prc23,metal-25},
$\beta$ delayed $\alpha$ emission from $^{16}$N~\cite{sa-epja21},
and $E1$ and $E2$ transitions of
$^{12}$C($\alpha$,$\gamma$)$^{16}$O reaction~\cite{sa-prc19,sa-cpc25,sa-npa25}.

In this work, we study 
elastic $n$-$^{16}$O scattering at
$0\le E_n \le 4$~MeV,
where $E_n$ is the neutron energy in the laboratory frame,
within the framework of an EFT. 
As we will discuss in the next section,
we introduce one excited and nineteen resonant states of $^{17}$O
as relevant degrees of freedom, 
and we employ composite fields for them. 
In energy regions close to resonant poles,
one needs to treat interactions non-perturbatively~\cite{hfvk-epja21}.
By employing composite fields, one can expand the terms
in the unitarity limits~\cite{kghvk-prl17}: 
those terms are naturally matched to the effective range
parameters~\cite{k-npb97,bs-npa01,ah-prc05,gk-26}. 
We also introduce a breakup channel, 
$n$-$^{16}$O, and assume that the contribution from 
$\alpha$-$^{13}$C channel is small. 
By using those fields, we construct an effective Lagrangian up to 
next-to-next-to-next-to-leading order. 
We first derive the expression of the $S$ matrices of elastic
$n$-$^{16}$O scattering for seven channels.
Parameters of the theory are fitted to an evaluated data set of ENDF/B-VIII.0
and compare the fitted values of the parameters with those 
in Ref.~\cite{twc-npa93}. 
We also discuss the implications of the fitted parameters
on the extrapolation 
of the $S$ factor of $^{13}$C($\alpha$,$n$)$^{16}$O to the stellar energies.  

The present work is organized as follows.
In Section 2, we discuss the overview of the construction of the theory,
and in Section 3, the effective Lagrangian for this work
is presented. 
In Section 4, the expressions of the $S$ factors, reaction amplitudes, and total cross section are displayed.
In Section 5, the numerical results of the parameter fit are reported,
and in Section 6, the results and discussions of this work are presented.
In the Appendix, an expression of the projection operator for the $f_{7/2}$ channel is displayed.

\vskip 2mm \noindent 
{\bf 2. Overview of the construction of the theory}

In the previous work~\cite{sa-26}, as mentioned, 
we studied the calculation of the $S$ factor 
of $^{13}$C($\alpha$,$n$)$^{16}$O at the energy up to 
$1$~MeV, where the energy is of $\alpha$-$^{13}$C system 
in the center-of-mass frame, 
by constructing an EFT. 
Parameters of the theory were fitted 
to the experimental data of the $S$ factor of 
$^{13}$C($\alpha$,$n$)$^{16}$O, 
reported from the LUNA and JUNA collaborations, 
in an energy range from 0.23~MeV to 1~MeV.  
In the study, we included three 
resonant states of $^{17}$O; 
$s_{1/2(2)}$, $f_{5/2(3)}$, $p_{3/2(3)}$ ($l_{j(i\textrm{\tiny th})}$) 
states at resonant energies, 
$E_x= 6.3623, 7.16424, 7.215$~MeV, 
respectively~\cite{twc-npa93}. 
Because the resonant energy of the $s_{1/2(2)}$ state is very close to 
the $\alpha$-$^{13}$C breakup energy of $^{17}$O, the resonant peak
of $s_{1/2(2)}$ is not covered by the experimental data of the $S$ 
factor. It leads to a major uncertainty in the extrapolation of the 
$S$ factor to the stellar energies. 

In the present work, we study elastic $n$-$^{16}$O scattering
at neutron energies, 
$0\le E_n \le 4$~MeV in laboratory frame.~\footnote{
	In this work, we employ three representations of the energies, 
	$E_n$, $E$, and $E_x$. 
	$E_n$ is the neutron energy in the laboratory frame,
	$E$ is the energy of $n$-$^{16}$O system in center-of-mass frame,
	and $E_x$ is the resonant energies of $^{17}$O. 
	We fix the energy range of $E_n$ as $0\le E_n\le 4$~MeV,
	which corresponds to $0\le E \le 3.7647$~MeV
and $4.1436 \le E_x \le 7.9084$~MeV, 
where $E = \frac{16}{17}E_n$ and $E_x = E + Q$:
$Q$ is the Q value of the open $n$-$^{16}$O channel of $^{17}$O,
$Q=4.1436$~MeV. }
In this energy range of $E_n$, one has two open channels,
$n$-$^{16}$O and $\alpha$-$^{13}$C, for $^{17}$O, where the breakup energy 
of the $\alpha$-$^{13}$C channel is $E_n = 2.2156$~MeV. 
In the present explorative study, 
we ignore the $\alpha$-$^{13}$C channel.
The contribution of the $\alpha$-$^{13}$C channel is suppressed 
due to the penetration factor of the Coulomb interaction between
$\alpha$ and $^{13}$C. 
One may notice other contributions to the elastic $n$-$^{16}$O scattering
such as 
the breakup to the $n$-$^{12}$C system of $^{13}$C 
and the excited states of $^{13}$C and $^{16}$O. 
Those contributions are open at higher energies than $E_n = 4$~MeV. 
We regard them as irrelevant degrees of freedom at high energy. 

For the scattering channels of the $n$-$^{16}$O system, 
because of the scattering states of spin-1/2 and spin-0 particles,  
we consider seven $lj$ channels for 
$l=s,p,d,f$ (or $l=0,1,2,3$)
and $j = l \pm 1/2$ (and $j\ge 0$);  
$lj = $ \{$s_{1/2}$, $p_{1/2}$, $p_{3/2}$, $d_{3/2}$, $d_{5/2}$, $f_{5/2}$, 
$f_{7/2}$\}.
For the bound and resonant states of $^{17}$O, we include
one excited state and nineteen resonant states of $^{17}$O 
in the present study. 
(See Table \ref{table;parameters}.)  
To fit the parameters of the theory to experimental data, 
we employ the total cross section 
of elastic $n$-$^{16}$O scattering from an evaluated data set of 
ENDF/B-VIII.0~\cite{betal-nds18}, 
because the experimental data,
the total cross sections and the phase shifts, 
of elastic $n$-$^{16}$O scattering 
were published more than fifty years ago, 
e.g., in Refs~\cite{fc-pr58,fjf-prc73}: 
it was not easy to find a data table of the quantities
from those publications.

Two parameters, the energy and width, are assigned to each of  
the resonant states, which correspond to the coefficients of the $p^0$ 
and $p^2$ terms at leading order and next-to-leading order
where $p$ is the magnitude of relative momentum of the $n$-$^{16}$O system
in center-of-mass frame. 
Sharp resonant states can be described well by those two parameters. 
We introduce the higher-order corrections of 
the $p^4$ and $p^6$ terms at next-to-next-leading order and 
next-to-next-to-next-leading order to the resonance states with large widths,
to describe the shapes of the broad resonant tails. 
When fitting the parameters, 
we do not include the parameters of the three sharp resonant states 
because of the large error bars in the evaluated data.
We fix the energies and widths of the other three resonant states at 
$E_n> 4$~MeV, using the values in Ref.~\cite{twc-npa93}, 
which are included as the background contributions from high energy. 
Thus, we have thirty-four parameters to fit the data, and one can find
that the ENDF compilation data are reproduced well 
by using the fitted parameters. 

\vskip 2mm \noindent 
{\bf 3. Effective Lagrangian}

The effective Lagrangian of this study may be written down as 
\bea
{\cal L} &=& 
{\cal L}_0 + 
{\cal L}_R + 
{\cal L}_{int}\,, 
\eea
where ${\cal L}_0$ is the Lagrangian for the kinetic term of the 
elementary neutron and $^{16}$O fields,
${\cal L}_R$ is that for 
the bound and resonant states of $^{17}$O in terms of the composite fields, 
and ${\cal L}_{int}$ is that to describe the interactions between the 
neutron and $^{16}$O fields and the composite fields of $^{17}$O. 

The Lagrangian ${\cal L}_0$ is standard. 
One has its expression as
\bea
{\cal L}_0 &=& 
\psi_n^\dagger \left( 
i\partial_0 
+ \frac{\nabla^2}{2m_n} 
\right)\psi_n 
+ \phi^\dagger_O \left(
i\partial_0
+ \frac{\nabla^2}{2m_O}
\right) \phi_O \,,
\eea
where $\psi_n$ is the nonrelativistic spin-1/2 neutron field,
and $\phi_O$ is the nonrelativistic scalar $^{16}$O filed. 
$m_n$ and $m_O$ are the masses of neutron and $^{16}$O, respectively. 

The Lagrangian ${\cal L}_R$ represents the bound and resonant states 
of $^{17}$O in terms of the composite fields, and 
one may have
\bea
{\cal L}_R &=& \sum_{lj(i)} \sum_{n=0}^N C_{lj(i)}^{(n)} 
d_{lj(i)}^\dagger \left(
i\partial_0
+ \frac{\nabla^2}{2m_{lj(i)}}
\right)^n d_{lj(i)}\,.
\eea
where $lj(i)$ denote $i$th states of the bound and resonant states
of $^{17}$O for $lj$ channels.
We have seven spin-angular momentum channels for $l=s,p,d,f$ and 
$j=l\pm 1/2$ (and $j\ge 0$) as 
\bea
lj(i) =\{ s_{1/2(i)}, 
p_{1/2(i)}, 
p_{3/2(i)}, 
d_{3/2(i)},
d_{5/2(i)},
f_{5/2(i)},
f_{7/2(i)}\}\,.
\eea
As displayed in Table \ref{table;parameters}, 
we include 
two states for $p_{1/2(i)}$ 
and $d_{5/2(i)}$ with $i=2,3$, 
three states for $s_{1/2(i)}$, 
$p_{3/2(i)}$, $f_{7/2(i)}$ with $i=1,2,3$, 
and $f_{5/2(i)}$ with $i= 2,3,4$, 
and four states for $d_{3/2(i)}$ with $i=1,2,3,4$
in the present study. 
The Lagrangian is expanded in terms of the number of derivatives on the fields.
The form, $i\partial_0 + \nabla^2/2m_{lj(i)}$, is required to be invariant 
under the Galilean transformation, and 
when one chooses the center-of-mass frame, the second term vanishes. 
We fix the number of the perturbation expansion terms as $N=1$, 
up to the next-leading order except for five states, 
\{$s_{1/2(1)}$, 
$s_{1/2(2)}$, 
$p_{3/2(3)}$,
$d_{3/2(1)}$, 
$d_{3/2(3)}$\}. 
$C_{lj(i)}^{(n)}$ with $n=0,1$ are the coefficients of the composite fields,
and they are fixed
by two effective range parameters, $a_0$ and $r_0$, or the resonant
energy and width, $E_R$ and $\Gamma_R$. 
In addition, we introduce higher-order terms; 
$C_{lj(i)}^{(n)}$ with $n=2$ for the $s_{1/2(1)}$ and $d_{3/2(1)}$ states,
which are fixed by an effective range parameter, $P_0$, and
a shape parameter, $a_{s_{1/2(1)}}$, respectively,
and $C_{lj(i)}^{(n)}$ with $n=2,3$ 
for the $s_{1/2(2)}$, $p_{3/2(3)}$ and $d_{3/2(3)}$ states,
which are fixed by two shape parameters, $a_{lj(i)}$ and $b_{lj(i)}$.
(The shape parameters, $a_{lj(i)}$ and $b_{lj(i)}$, are defined
in Eq.~(\ref{eq;anb}).)
$d_{lj(i)}$ are the composite fields for 
the bound and resonant states of $^{17}$O. 
$d_{s_{1/2(i)}}$ and $d_{p_{1/2(i)}}$ 
are two spinors for $j=1/2$, 
$d_{p_{3/2(i)}}$ and $d_{d_{3/2(i)}}$
are four spinors for $j=3/2$, 
$d_{d_{5/2(i)}}$ and $d_{f_{5/2(i)}}$ are six spinors for $j=5/2$,
and $d_{f_{7/2(i)}}$ are an eight spinor for $j=7/2$. 

The Lagrangian ${\cal L}_{int}$ to connect the $n$-$^{16}$O states and 
the excited and resonant states of $^{17}$O may be obtained as
\bea
{\cal L}_{int} &=& -\sum_{lj(i)} y_{lj(i)}
\left\{d_{lj(i)}^\dagger \left[P_{lj}(\psi_n O_l \phi_O)\right]
+ \left[P_{lj}(\psi_n O_l \phi_O)\right]^\dagger d_{lj(i)}
	\right\}\,,
\eea
where $y_{lj(i)}$ are the coupling constants between the $n$-$^{16}$O states
and the bound and resonant states of $^{17}$O. 
The indices of the angular momentum projection operators $O_l$ and 
the projection operators $P_{lj}$ are suppressed in the above expression. 
Those coupling constants are redundant 
(they appear in the form, $C_{lj(i)}/y_{lj(i)}^2$, 
in the scattering amplitudes) and
are rewritten, along with the coefficients, $C_{lj(i)}$, 
as the effective range parameters or the resonant energy and width. 
$P_{lj}$ are the projection operators 
in $2\times 2$, $2\times 4$, $2\times 6$,
and $2\times 8$ matrix forms to connect the spin-1/2 neutron field and 
two, four, six, and eight spinors of $d_{lj(i)}$ fields for 
$j=1/2,3/2,5/2,7/2$, respectively. 
$O_l$ are the projection operators of the $n$-$^{16}$O states in the 
relative angular momentum states for $l=0,1,2,3$.
The expression of the projection operator $P_{lj}$ for $lj=f_{7/2}$ is 
presented in the Appendix.  
The projection operators
$P_{lj}$ for $l=s,p,d,f$ and $j=1/2,3/2,5/2$ 
can be found 
in Appendix A in Ref.~\cite{sa-26},
and the projection operators $O_l$ for $l=0,1,2,3$ can be found in 
Appendix B in Ref.~\cite{sa-26}. 

\vskip 2mm \noindent
{\bf 4. $S$ matrices, reaction amplitudes, and total cross section}

The $S$ matrices for the $lj$ channels are related to phase shifts 
and scattering amplitudes as
\bea
S_{lj} &=& e^{2i\delta_{lj}}
= 1 +2ip \tilde{A}_{lj} \,,
\eea
where $S_{lj}$ are the $S$ matrices, 
$\delta_{lj}$ are the phase shifts, and 
$\tilde{A}_{lj}$ are the scattering amplitudes
for the $lj$ channels.
As mentioned, 
$p$ is the magnitude of relative momentum of the $n$-$^{16}$O system
in center-of-mass frame. 

\begin{figure}
\begin{center}
  \includegraphics[width=3.5cm]{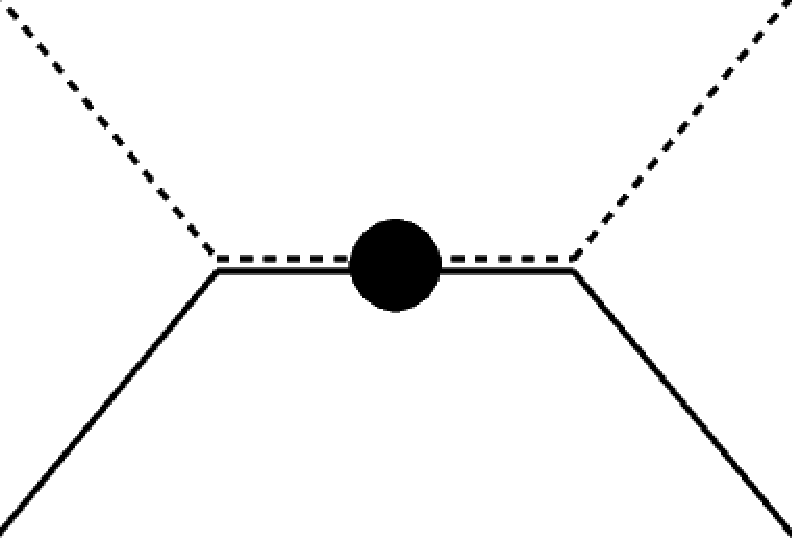}
\caption{
	Feynman diagram of elastic $n$-$^{16}$O scattering.
	Solid lines denote incoming and outgoing neutrons, 
dashed lines do incoming and outgoing $^{16}$Os.
A solid-and-dashed-double line with a filled circle denotes the 
	dressed propagator of the bound and resonant states of $^{17}$O.
	See Fig.~\ref{fig;propagator} as well. 
}
\label{fig;amplitudes}       
\end{center}
\end{figure}

\begin{figure}
\begin{center}
  \includegraphics[width=10.5cm]{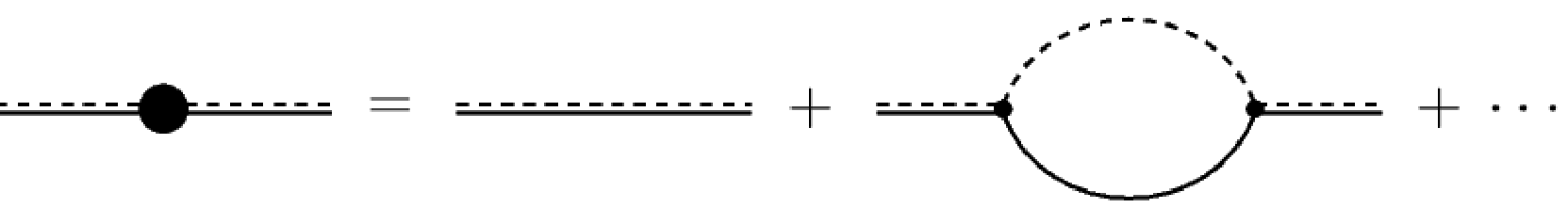}
\caption{
	Diagrams for the dressed propagator 
	of the bound and resonant states of $^{17}$O.
A solid line represents a neutron propagator,
	and a dashed line is a propagator of $^{16}$O. 
	A solid-and-dashed-double line with and without a filled circle
	represent a dressed and bare propagator 
	of the states of $^{17}$O, respectively.
}
\label{fig;propagator}       
\end{center}
\end{figure}

We now assume that the phase shifts, $\delta_{lj}$, 
are decomposed as~\cite{sa-prc23,sa-prc22}
\bea
\delta_{lj} &=& 
\delta_{lj(1)} 
+ \delta_{lj(2)} 
+ \delta_{lj(3)} 
+ \cdots \,,
\eea
where $\delta_{lj(1)}$,
$\delta_{lj(2)}$ and
$\delta_{lj(3)}$ 
are the phase shifts 
generated from, e.g., 
the first, second, and third resonant states, respectively.
Those phase shifts are related to the scattering amplitudes 
of the resonant states as
\bea
e^{2i\delta_{lj(i)}} &=& 1 + 2ip\tilde{A}_{lj(i)}\,,
\label{eq;exp2idellji}
\eea
where $\delta_{lj(i)}$ are the phase shifts 
and $\tilde{A}_{lj(i)}$ are the scattering amplitudes 
generated from the $i$th resonant states for the $lj$ channels.
The scattering amplitudes, $\tilde{A}_{lj(i)}$,  
are calculated from the Feynman diagrams
depicted in Figs.~\ref{fig;amplitudes} and \ref{fig;propagator}. 

The expression of the scattering amplitude 
from the first excited $s_{1/2(1)}$ state of $^{17}$O 
is standard~\cite{ah-prc05} 
and is obtained as 
\bea
\tilde{A}_{s_{1/2(1)}} &=& \frac{1}{K_0(p) -ip}\,,
\label{eq;As1/2(1)}
\eea
where the $-ip$ term is generated from the bubble diagram in 
Fig.~\ref{fig;propagator}, due to the propagation of neutron and $^{16}$O,
and we include three terms of $C_{s_{1/2(1)}}^{(n)}$ with $n=0,1,2$ for 
the composite field for the $s_{1/2(1)}$ state 
of $^{17}$O.~\footnote{ 
Infinities from the loop integral are
subtracted by the coefficients, $C_{lj(i)}^{(n)}$. 
}
The three terms are represented 
in terms of effective range parameters 
in the function $K_0(p)$ 
as 
\bea
K_0(p) &=& -\frac{1}{a_0} + \frac12r_0p^2 
- \frac14 P_0p^4\,,
\eea
where $a_0$, $r_0$, $P_0$ are scattering length, effective range, 
shape parameter, respectively. 

We may fix one of the effective range parameters by using the condition 
that the inverse of the amplitude vanishes 
at the binding energy, i.e.,
$K_0(p=i\gamma_0) + \gamma_0 = 0$ at $E=-B_0$ (or $p =  i \gamma_0$), where
$B_0$ is the binding energy of the $n$-$^{16}$O system
($\gamma_0$ is the binding momentum, $\gamma_0=\sqrt{2\mu B_0}$,
and $\mu$ is the reduced mass of $n$ and $^{16}$O) 
of the first excited $s_{1/2}$ state of $^{17}$O at $E_x=0.870756$~MeV; 
$B_0 = Q - E_x=3.27284$~MeV where $Q$ is the Q value of the $n$-$^{16}$O
open channel of $^{17}$O, $Q=4.1436$~MeV. 
Thus, we rewrite the function of $K_0(p)$ as
\bea
K_0(p) = -\gamma_0 + \frac12 r_0(\gamma_0^2 + p^2)
+ \frac14 P_0(\gamma_0^4 - p^4)\,,
\eea
where 
$\gamma_0 = 76.0603$~MeV. 
In addition, we may restrict the effective range, $r_0$, by using the data of the 
total cross section of the 
elastic scattering at the thermal neutron energy ($v=2200$~m/s),
$\sigma_{th} = 3.89421$~b~\cite{betal-nds18}. 
One may have a relation,
\bea
\sigma_{th} = \frac{4\pi}{
	\left(
	-\gamma_0 + \frac12r_0\gamma_0^2
	+ \frac14P_0\gamma_0^4
	\right)^2
}\,,
\eea
where we have assumed zero momentum, $p=0$, and have
\bea
r_0 &=& \frac{2}{\gamma_0}\left(
1 - \frac14 P_0\gamma_0^3 
-\frac{2}{\gamma_0}\sqrt{
	\frac{\pi}{\sigma_{th}}
}
\right)\,,
\eea
where $P_0$ is fitted to the experimental data of the elastic scattering. 

The expression of the scattering amplitudes from the $i$th resonant states
for the $lj$ channels, 
including four coupling constants,
$C_{lj(i)}^{(n)}$ with $n=0,1,2,3$, are obtained as~\cite{sa-prc23} 
\bea
\tilde{A}_{lj(i)} &=&  
- \frac{1}{p}
\frac{\frac12 \Gamma_{lj(i)}(E)
	}{ 
	E - E_{Rlj(i)} + R_{lj(i)}(E)+  i \frac12\Gamma_{lj(i)}(E)
	} \,,
	\label{eq;Alj(i)}
\eea
where the denominator of the real part of the amplitude 
is expanded around $E=E_{Rlj(i)}$, 
and we include the terms up to the $E^3$ order
(the $p^6$ order, where $E=p^2/(2\mu)$).
One has
\bea
\Gamma_{lj(i)}(E) &=& \Gamma_{Rlj(i)} \left(
\frac{E}{E_{Rlj(i)}}
\right)^{(2l+1)/2}\,,
\\
R_{lj(i)}(E) &=& a_{lj(i)}(E - E_{Rlj(i)})^2 
+ b_{lj(i)}(E - E_{Rlj(i)})^3 \,,
\label{eq;anb}
\eea
where $E_{Rlj(i)}$ and $\Gamma_{Rlj(i)}$ are the resonant energy
and width of the $i$th resonant states for the $lj$ channels.
The functions, $R_{lj(i)}(E)$, contain the second and third order terms
expanded around $E=E_{Rlj(i)}$;
the coefficients $a_{lj(i)}$ and $b_{lj(i)}$ are fitted to 
the experimental data of 
the shapes of the broad tails of the resonant states with large widths.   

Using Eqs.~(\ref{eq;exp2idellji}), (\ref{eq;As1/2(1)}), (\ref{eq;Alj(i)}),
one has a simple and transparent expression 
of the $S$ matrices~\cite{sa-prc23,sa-prc22}.
The $S$ matrix for the $s_{1/2}$ channel, in which we have
the first excited $s_{1/2(1)}$ state of $^{17}$O, is obtained as
\bea
S_{s_{1/2}} &=& e^{2i\delta_{s_{1/2}}} 
= \frac{K_0(p) + ip}{K_0(p) -ip}
\prod_i
\frac{E-E_{Rs_{1/2(i)}} + R_{s_{1/2(i)}}(E) - i \frac12\Gamma_{s_{1/2(i)}}(E)}{
E - E_{Rs_{1/2(i)}} + R_{s_{1/2(i)}}(E) + i \frac12\Gamma_{s_{1/2(i)}}(E) 
}\,,
\eea
where $i=2,3$. 
In addition, the $S$ matrices for the other $lj$ channels are obtained as 
\bea
S_{lj} &=& e^{2i\delta_{lj}}  = 
\prod_i \frac{E-E_{Rlj(i)} + R_{lj(i)}(E) - i \frac12\Gamma_{lj(i)}(E)}{
	E - E_{Rlj(i)} + R_{lj(i)}(E) + i \frac12\Gamma_{lj(i)}(E) }
	\,,
\eea
where the $i$th resonant states 
for the $lj$ channels for this study 
are listed in Table \ref{table;parameters}. 

The expression of the total cross section of elastic scattering 
for spin-1/2 and spin-0 particles in terms of the phase shifts 
is well known~\cite{cd-pr49}. 
Using Eqs.~(2.399) to (2.404) in Ref.~\cite{f-13}, one has
\bea
\sigma(E) &=& \frac{4\pi}{p^2}\left[
\sum_{l=0}^3 (2l+1) |f_l|^2 
+ \sum_{l=1}^3 \frac{l(l+1)}{2l+1}|g_l|^2
\right]\,,
\eea
where
\bea
f_l &=&
\frac{1}{2i}
\left[
	\frac{l+1}{2l+1}\left(
	e^{2i\delta_l^+} -1
	\right)
	+ \frac{l}{2l+1}\left(
	e^{2i\delta_l^-} -1
	\right)
	\right] \,,
\\
g_l &=& 
\frac{1}{2i}\left(
e^{2i\delta_l^+}
- e^{2i\delta_l^-}
\right)\,,
\eea
where we employed a different notation of the phase shifts, $\delta_l^\pm$,
which correspond to $\delta_{lj}$ with $j=l\pm 1/2$ and $j\ge 0$.

\vskip 2mm \noindent
{\bf 5. Numerical results}

\begin{figure}[t]
\begin{center}
	  \includegraphics[width=11cm]{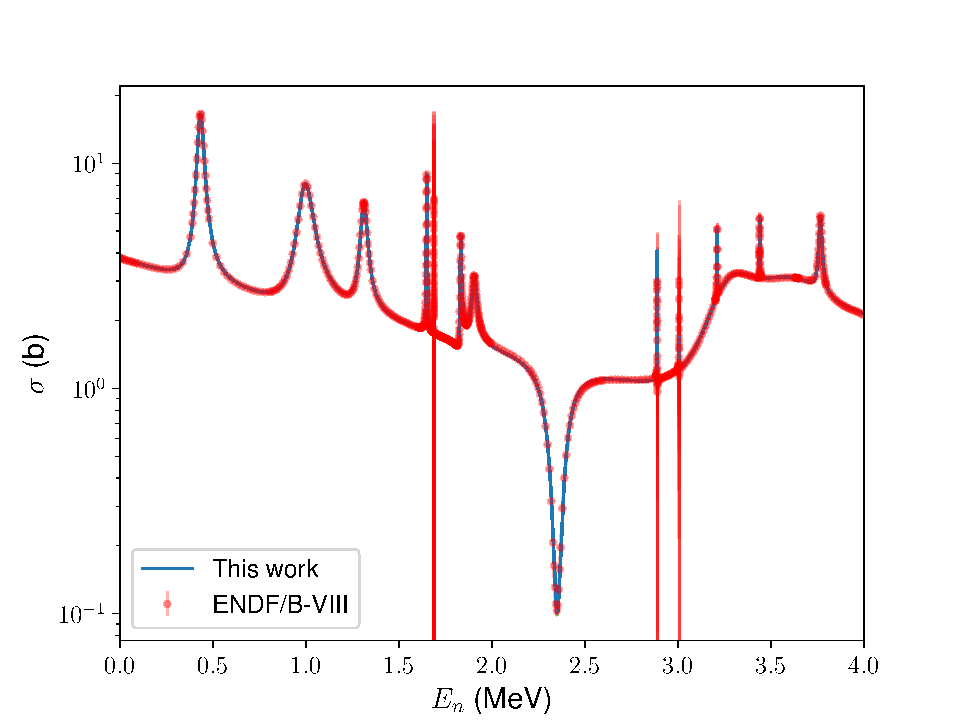}
	\caption{Total cross section of elastic $n$-$^{16}$O scattering
	as a function of neutron energy, $E_n$, in laboratory frame.  
	A line of this work is plotted by using the fitted 
	value of the parameters in Table \ref{table;parameters}.
	Evaluated data of ENDF/B-VIII.0 are displayed in the figure as well.
}
\label{fig;cross_section}       
\end{center}
\end{figure}

Among the four bound states 
(one ground state and three excited states) of $^{17}$O, 
\{$d_{3/2(1)}$, $s_{1/2(1)}$, $p_{1/2(1)}$, $f_{5/2(1)}$\},
we include the first excited $s_{1/2(1)}$ state in this study.  
As mentioned, we employ the evaluated data of the total cross section 
of elastic $n$-$^{16}$O scattering at 
$0\le E_n \le 4$~MeV from ENDF/B-VIII.0~\cite{betal-nds18}.
The data are displayed in Fig.~\ref{fig;cross_section}. 
We include sixteen resonant states of $^{17}$O in this energy range.
(See Table \ref{table;parameters}.)
We also include three resonant states with large widths, 
\{$s_{1/2(3)}$, $p_{1/2(3)}$, $d_{3/2(4)}$\}, 
as background contributions from high energy, 
whose resonant energies are larger than $E_n = 4$~MeV. 
To fit the parameters, we perform a $\chi^2$ fit in an 
MCMC analysis by employing the \texttt{emcee} package~\cite{emcee}.

\begin{table}
\begin{center}
\begin{tabular}{c|cc|cccc}
	$l_{j^\pi}$($i$th) & $E_x$ (MeV)\cite{twc-npa93} & 
	$\Gamma$ (keV)\cite{twc-npa93} &
	$E_x$ (MeV) & $\Gamma$ (keV) & 
	$a$ (MeV$^{-1}$) & $b$ (MeV$^{-2}$) \cr
		\hline 
	$s_{1/2^+(1)}$ & 0.870756(20) & & 
		0.870756$^*$ & & $-5.73(6)^{**}$ & \cr
	$p_{3/2^-(1)}$ & 4.5517(7) & 38.7(28) & 
		4.5453(1) & 41.2(2) & & \cr
	$d_{3/2^+(1)}$ & 5.0868(9) & 90(3) &
	5.0846(3) & 90.7(5) & 0.15(3) & \cr
	$p_{3/2^-(2)}$ & 5.3870(22) & 37.1(24) &
	5.3788(3) & 39.4(6) & & \cr
	$f_{7/2^-(1)}$ & 5.6973(33) & 3.4(3) &
	5.6970(1) & 3.7(1) & & \cr
	$f_{5/2^-(2)}$ & 5.73204(40) & $<1$ &
	[5.73191(5) & 0.103(5)] & & \cr
	$d_{3/2^+(2)}$ & 5.86962(40) & 6.6(7) &
	5.8681(2) & 7.2(2) & & \cr
	$p_{1/2^-(2)}$ & 5.9315(18) & 32(3) &
	5.9391(7) & 28(1) & & \cr
	$s_{1/2^+(2)}$ & 6.3623(29) & 126(14) &
	6.3593(7) & 125.5(19) & $-0.31(7)$ & 0.045(30) \cr
	$d_{5/2^+(2)}$ & 6.8606(4) & $<1$ &
	[6.86272(3) & 0.103(5)] & & \cr
	$f_{7/2^-(2)}$ & 6.9725(4) & $<1$ &
	[6.97237(10) & 0.30(11)] & & \cr
	$f_{5/2^-(3)}$ & 7.16424(17) & 1.38(5) &
	7.1664(1) & 1.7(1) & & \cr
	$d_{3/2^+(3)}$ & 7.215(5) & 263(7) &
	7.2112(25) & 300(6) & 0.49(11) & 0.61(10) \cr
	$d_{5/2^+(3)}$ & 7.37752(19) & $0.61^{+14}_{-11}$ &
	7.38032(8) & 0.49(11) & & \cr
	$f_{5/2^-(4)}$ & 7.38064(14) & $0.90^{+17}_{-14}$ & 
	7.38268(12) & 1.3(2) & & \cr
	$p_{3/2^-(3)}$ & 7.542(20) & 500(5) &
	7.5290(28) & 551(7) & 0.66(2) & 0.15(1) \cr
	$f_{7/2^-(3)}$ & 7.68732(22) & 14.4(3) &
	7.68879(23) & 17.5(5) & & \cr
	$s_{1/2^+(3)}$ & 7.954(8) & 85(9) &
	 & & & \cr
	$p_{1/2^-(3)}$ & 7.99(5) & 270(27) &
	& & & \cr
	$d_{3/2^+(4)}$ & 8.068(10) & 77(8) &
	& & & \cr
	\hline
\end{tabular}
	\caption{
		Fitted values of energy and width of 
		$i$th states of $lj$ channels of $^{17}$O 
		including shape parameters, 
		$a_{lj(i)}$ and $b_{lj(i)}$,
		where the labels $lj(i)$ are suppressed in the table. 
		Resonant energies and widths in the second and third
		columns are values 
		from the TUNL nuclear data compilation~\cite{twc-npa93} for 
		comparison. 
		Values in the third to sixth columns are the fitted 
		values of the present work. 
		Three states at the bottom of the table are included as
		contributions from high energy, and those parameters 
		are not fitted to the data. 
		Values of $a$ and $b$  
		in fifth and sixth columns are higher-order terms 
		which are fitted to the shapes of broad tails 
		of the resonances with large widths. 
		$^*$It is used to fix a value of $\gamma_0$.
		$^{**}$A fitted value of an effective range
		parameter, $P_0$ (fm$^3$). 
		Values in the square brackets are fitted in 
		the first step of the fit. 
        See the text as well. 
	}
	\label{table;parameters}
\end{center}
\end{table}

When fitting the parameters, 
because large error bars appear in three sharp resonant $f_{5/2(2)}$,
$d_{5/2(2)}$, $f_{7/2(2)}$ states in Fig.~\ref{fig;cross_section}, 
we work out the fit in two steps.
In the first step, we include the parameters up to the next-leading order
(at the $p^0$ and $p^2$ orders), 
the resonant energies and widths. The number of parameters is 
thirty-two, where the effective range parameter $r_0$ 
is fixed without including $P_0$, $r_0 =2.77055$~fm. 
The $\chi^2$ value of the fit is $\chi^2/N = 9.44$ 
where $N$ is the number of data points, $N=742$. 
In the second step, we fix the six parameters of the three sharp resonant
states with the large error bars 
by using the parameters fitted in the first step. 
The fitted values of the 
parameters in the first step are displayed in the square brackets 
in Table \ref{table;parameters}. 
Then, we include eight parameters at the $p^4$ and $p^6$ orders 
in the first excited $s_{1/2(1)}$ state and four resonant $d_{3/2(1)}$, 
$s_{1/2(2)}$, $d_{3/2(3)}$, $p_{3/2(3)}$ states, which have relatively
large widths; 
we now have thirty-four parameters.  
The $\chi^2$ value of the fit in the second step is $\chi^2/N=0.268$, 
and fitted values of the parameters are listed in Table \ref{table;parameters}. 
For comparison, values of the energies and widths of the resonant states 
of $^{17}$O in the TUNL nuclear data compilation~\cite{twc-npa93} 
are also displayed in the table. 
In Fig.~\ref{fig;cross_section}, 
we plot a line of the total cross section as a function of the 
neutron energy, $E_n$, by using the fitted values of the parameters,
and find that the evaluated data of ENDF/B-VIII.0
is reproduced well.

\begin{figure}[t]
\begin{center}
	  \includegraphics[width=12cm]{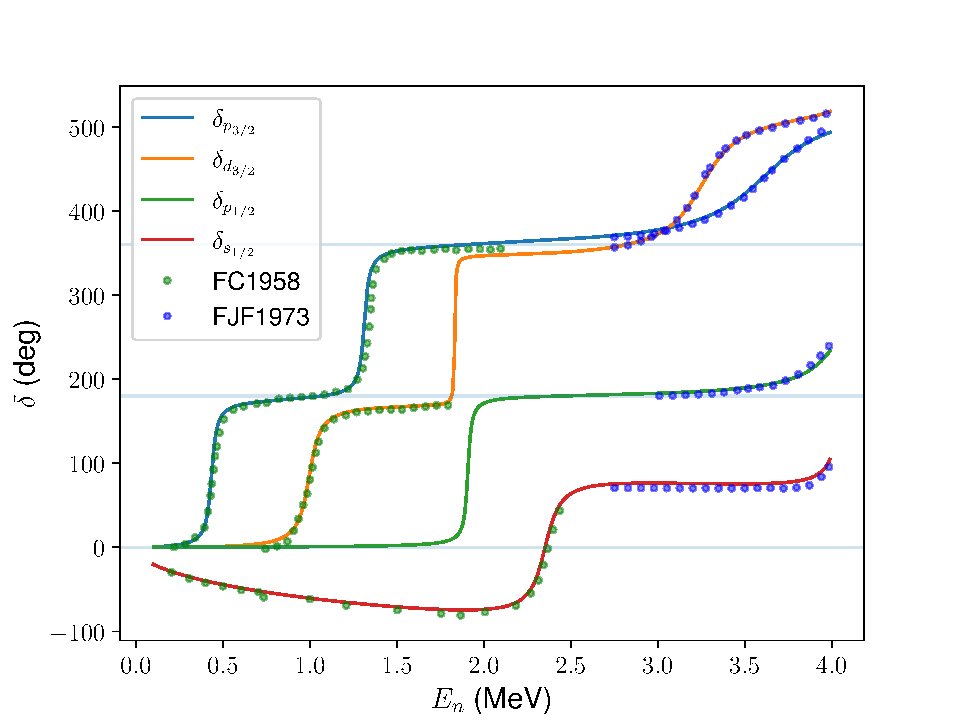}
\caption{
	Phase shifts of elastic $n$-$^{16}$O scattering for 
	$s_{1/2}$, $p_{1/2}$, $p_{3/2}$, $d_{3/2}$ channels as functions of 
	neutron energy, $E_n$, are plotted by using the fitted parameters
	in Table~\ref{table;parameters}. 
	Phase shift data from FC1958~\cite{fc-pr58} 
	and FJF1973~\cite{fjf-prc73} are
	displayed in the figure as well. 
}
\label{fig;phase_shifts}       
\end{center}
\end{figure}
In Fig.~\ref{fig;phase_shifts}, we plot the phase shifts of 
the elastic $n$-$^{16}$O scattering for $s_{1/2}$, $p_{1/2}$, $p_{3/2}$,
$d_{3/2}$ channels as functions of the neutron energy, $E_n$, by using
the fitted values of the parameters in Table~\ref{table;parameters}. 
The phase shift data in Fowler and Cohn (FC1958)~\cite{fc-pr58} and
Fowler, Jonson, and Feezel (FJF1973)~\cite{fjf-prc73} are included in the same figure.\footnote{
	The phase shift data are read from the figures 
	in Refs.~\cite{fc-pr58,fjf-prc73}
by using a utility WebPlotDigitizer~\cite{wpd}.}
One can find that the plotted lines reproduce those phase shift data well.

\vskip 2mm \noindent
{\bf 6. Results and discussion}

In this work, the elastic $n$-$^{16}$O scattering at the neutron energy,
$0\le E_n \le 4$~MeV, for the seven channels, 
$lj = \{s_{1/2}, p_{1/2}, p_{3/2}, d_{3/2}, d_{5/2}, f_{5/2}, f_{7/2} \}$, 
is studied within the framework of an EFT. 
We included one excited state and nineteen resonant states of $^{17}$O,
as relevant degrees of freedom for the present study, 
and derive the expression of the $S$ matrices of the elastic scattering 
for the seven channels from the effective Lagrangian. 
Thirty-four parameters of the theory are fitted to the evaluated data 
of ENDF/B-VIII.0, and we find that the evaluated data are well reproduced
by using the fitted values of the parameters. 
We note that with the parameters in the first step of the fit,
the fitted line can reproduce the data of the resonances with small widths,
but the discrepancies between the plotted line and the data appear
at the regions between the resonances with large widths.
(We did not show it in the figure.) 
Those discrepancies due to the broad resonant tails are 
completely fitted by the shape parameters, $a_{lj(i)}$ and $b_{lj(i)}$, 
in the second step of the fit, 
and the whole range of the data is reproduced well.
We also plot the phase shifts for the $s_{1/2}$, $p_{1/2}$, $p_{3/2}$, 
$d_{3/2}$ channels and find that the plotted lines well reproduce
the phase shift data in the literature. 

The evaluated data of the elastic $n$-$^{16}$O scattering at 
the neutron energy, $0\le E_n \le 8$~MeV in ENDF/B-VIII.0 
are calculated by an $R$-matrix analysis~\cite{betal-nds18}.
Our result indicates that the formalism of EFT can be an alternative framework,
along with the $R$-matrix analysis, 
for the study of elastic $n$-$^{16}$O scattering at low energies. 
Notable differences between the EFT and the $R$-matrix analysis are
as follows: 
1) two parameters, a matching radius $a$ and a boundary parameter $B_c$ 
in the $R$ matrix analysis~\cite{sl-epjwc23} do not exist in the EFT,
2) the expression of the $S$ matrices in the EFT is simple and transparent,
and the energies and widths of the resonant states appearing in the data 
are directly fitted by the parameters of the EFT, 
and 3) the EFT needs to introduce the shape parameters, 
$a_{lj(i)}$ and $b_{lj(i)}$, for the resonances with large widths:
those shape parameters do not exist 
in the $R$-matrix theory~\cite{lt-rmp58}. 
Because of those differences, 
even though the two methods can equally well fit the same data set,
the values of the energies and widths of the resonant states can be
different; 
a model dependence (or a systematic uncertainty)
between the two methods may appear
from the resonance states with large widths. 
As seen in Table~\ref{table;parameters}, 
the four resonant $s_{1/2(2)}$, $p_{3/2(3)}$, $d_{3/2(1)}$, $d_{3/2(3)}$ 
states, to which we needed to introduce the shape parameters,  
have the widths larger than  $\Gamma = 90$~keV.

Especially, the resonant $s_{1/2(2)}$ and $d_{3/2(3)}$ 
states of $^{17}$O may bring the uncertainties in the 
extrapolation of the $S$ factor of $^{13}$C($\alpha$,$n$)$^{16}$O
to the stellar energies. 
One may find some scattered values of the resonant energies and widths
of those states in Table 1 in Ref.~\cite{cetal-jpg24} 
and Table I in Ref.~\cite{fmhw-prc15}. 
Thus, one should be cautious when choosing one of those evaluated values 
of the parameters for his or her study. 
To avoid such a systematic uncertainty, one may need to analyze simultaneously 
the two data sets, the elastic $n$-$^{16}$O scattering 
and the $^{13}$C($\alpha$,$n$)$^{16}$O reaction, 
for the study of the $S$ factor of $^{13}$C($\alpha$,$n$)$^{16}$O 
at the stellar energies.

\vskip 2mm \noindent
{\bf Acknowledgements}

The author would like to thank 
Seung-Woo Hong for the useful information 
about the covariance matrix, File 33, of ENDF/B-VIII.0.  
This work was supported by the National Research Foundation grant, funded by the Korean government's Ministry of Science and ICT 
(Grant No. RS-2025-16065411).

\vskip 2mm \noindent
{\bf Appendix: Projection operator}

In this Appendix, we present the expression of the projection operator
$P_{lj}$ for $lj=f_{7/2}$. Those of the other projection operators $P_{lj}$
for $j=1/2,3/2,5/2$ can be found in Appendix A in Ref.~\cite{sa-26}.
This term is presented in the interaction Lagrangian as
\bea
d_{f_{7/2}}^\dagger \left[ P_{f_{7/2},ijk} 
\left(\psi_n O_{ijk}^{(l=3)}\phi_O
\right)
\right]\,,
\eea
where the $P_{f_{7/2}}$ operator is presented as a $2\times 8$ matrix
between an eight spinor of $d_{f_{7/2}}^\dagger$ and a two spinor of 
$\psi_n$. $O_{ijk}^{(l=3)}$ is presented in terms of Cartesian tensors.
The $f_{7/2}$ state from $l=3$ and $s=1/2$ is given by the spin-angular
function in Eq.~(3.7.64) in Ref.~\cite{sakurai} as 
\bea
\frac{\chi^\dagger}{\sqrt7}
\left(
\begin{tabular}{cccccccc}
	$\sqrt7 T_{33}$ & $\sqrt6 T_{32}$ & $\sqrt5 T_{31}$ & $\sqrt4 T_{30}$ &
	$\sqrt3 T_{3-1}$ & $\sqrt2 T_{3-2}$ & $T_{3-3}$ & 0 \cr
	0 & $T_{33}$ & $\sqrt2 T_{32}$ & $\sqrt3 T_{31}$ & $\sqrt4 T_{30}$ &
	$\sqrt5 T_{3-1}$ & $\sqrt6 T_{3-2}$ & $\sqrt7 T_{3-3}$ 
\end{tabular}
\right)d_{f_{7/2}}\,,
\label{eq;f3/2}
\eea
where $T_{ij}$ are the Cartesian tensor for $l=3$.  
The expression of $T_{ij}$ may be given as
\bea
T_{3\pm3} &=& l_\pm l_\pm l_\pm\,,
\\
T_{3\pm2} &=& \frac{1}{\sqrt3}\left(
l_\pm l_\pm l_0 + l_\pm l_0 l_\pm + l_0 l_\pm l_\pm
\right)\,,
\\
T_{3\pm1} &=& \frac{1}{\sqrt{15}}\left(
2 l_\pm l_0 l_0 + 2 l_0 l_\pm l_0 + 2 l_0 l_0 l_\pm 
+ l_\pm l_\pm l_\mp + l_\pm l_\mp l_\pm + l_\mp l_\pm l_\pm
\right)\,,
\\
T_{30} &=& \sqrt{\frac25}\left[
	l_0l_0l_0 + \frac12\left(
	l_+l_-l_0 + l_+ l_0l_- + l_0 l_+l_- + l_0l_-l_+ 
	+ l_-l_+l_0 + l_-l_0l_+
	\right)
	\right]\,,
\eea
where $\vec{l}$ is the relative momentum between neutron and $^{16}$O,
and
\bea
l_\pm = \mp \frac{1}{\sqrt2} (l_1 \pm i l_2)\,, \ \ \ 
l_0 = l_3\,.
\eea
Now the matrix in Eq.~(\ref{eq;f3/2}), $M$, is decomposed as
\bea
M &=& LCR\,,
\eea
with
\bea
L &=& \left(
\begin{tabular}{cccc}
	$l_+$ & $\sqrt2 l_0$ & $l_-$ & 0 \cr
	0 & $l_+$ & $\sqrt2 l_0$ & $l_-$ 
\end{tabular}
\right)\,,
\\
C &=& \left(
\begin{tabular}{cccccc}
	$l_+$ & $\sqrt2 l_0$ & $l_-$ & & & \cr
	& $l_+$ & $\sqrt2 l_0$ & $l_-$ & &  \cr
	& & $l_+$ & $\sqrt2 l_0$ & $l_-$ &   \cr
	& & & $l_+$ & $\sqrt2 l_0$ & $l_-$    \cr
\end{tabular}
\right)\,,
\\
R &=& \left(
\begin{tabular}{cccccccc}
	$\sqrt7 l_+$ & $\sqrt2 l_0$ & $1/\sqrt3 l_-$ & & & & & \cr
	& $l_+$  & $\sqrt{2/3} l_0$ & $1/\sqrt5 l_-$ & & & & \cr
	& & $1/\sqrt3 l_+$  & $\sqrt{2/5} l_0$ & $1/\sqrt5 l_-$ & & & \cr
	& & & $1/\sqrt5 l_+$  & $\sqrt{2/5} l_0$ & $1/\sqrt3 l_-$ & & \cr
	& & & & $1/\sqrt5 l_+$  & $\sqrt{2/3} l_0$ & $l_-$ & \cr
	& & & & & $1/\sqrt3 l_+$  & $\sqrt{2} l_0$ & $\sqrt7 l_-$ \cr
\end{tabular}
\right)\,,
\eea
where the blank elements in the above matrices are zero. 

The matrices, $L$, $C$, $R$, and further decomposed in a vector form as 
\bea
L = \vec{L}\cdot \vec{l}\,, \ \  
C = \vec{C} \cdot \vec{l}\,, \ \ 
R = \vec{R}\cdot \vec{l}\,,
\eea
where 
\bea
L_1 &=& 
\frac{1}{\sqrt2} 
\left(
\begin{tabular}{cccc}
	$-1$ & 0 & 1 & 0 \cr
	0 & $-1$ & 0 & 1
\end{tabular}
\right)\,,
\\
L_2 &=& 
-\frac{i}{\sqrt2} 
\left(
\begin{tabular}{cccc}
	$1$ & 0 & 1 & 0 \cr
	0 & $1$ & 0 & 1
\end{tabular}
\right)\,,
\nnb 
\\
L_3 &=& 
\sqrt2 
\left(
\begin{tabular}{cccc}
	0 & 1 & 0 & 0 \cr
	0 & 0 & 1 & 0
\end{tabular}
\right)\,,
\\
C_1 &=& \frac{1}{\sqrt2}
\left(
\begin{tabular}{cccccc}
	$-1$ & 0 & 1 & & & \cr
& $-1$ & 0 & 1 & & \cr
& & $-1$ & 0 & 1 & \cr
& & & $-1$ & 0 & 1 \cr
\end{tabular}
\right)\,,
\\
C_2 &=& - \frac{i}{\sqrt2}
\left(
\begin{tabular}{cccccc}
	$1$ & 0 & 1 & & & \cr
& $1$ & 0 & 1 & & \cr
& & $1$ & 0 & 1 & \cr
& & & $1$ & 0 & 1 \cr
\end{tabular}
\right)\,,
\\
C_3 &=& \sqrt2
\left(
\begin{tabular}{cccccc}
	0 & 1 & 0 & & & \cr
& 0 & 1 & 0 & & \cr
& & 0 & 1 & 0 & \cr
& & & 0 & 1 & 0 \cr
\end{tabular}
\right)\,,
\\
R_1 &=& \frac{1}{\sqrt2}\left(
\begin{tabular}{cccccccc}
	$-\sqrt7$ & 0 & $1/\sqrt3$ & & & & & \cr
	& $-1$ & 0 & $1/\sqrt5$ & & & & \cr
	& & $-1/\sqrt3$ & 0 & $1/\sqrt5$ & & & \cr
	& & & $-1/\sqrt5$ & 0 & $1/\sqrt3$ & & \cr
	& & & & $-1/\sqrt5$ & 0 & $1$ & \cr
	& & & & & $-1/\sqrt3$ & 0 & $\sqrt7$ \cr
\end{tabular}
\right)\,,
\\
R_2 &=& - \frac{i}{\sqrt2}\left(
\begin{tabular}{cccccccc}
	$\sqrt7$ & 0 & $1/\sqrt3$ & & & & & \cr
	& $1$ & 0 & $1/\sqrt5$ & & & & \cr
	& & $1/\sqrt3$ & 0 & $1/\sqrt5$ & & & \cr
	& & & $1/\sqrt5$ & 0 & $1/\sqrt3$ & & \cr
	& & & & $1/\sqrt5$ & 0 & $1$ & \cr
	& & & & & $1/\sqrt3$ & 0 & $\sqrt7$ \cr
\end{tabular}
\right)\,,
\\
R_3 &=& \sqrt2\left(
\begin{tabular}{cccccccc}
	$0$ & 1 & $0$ & & & & & \cr
	& $0$ & $1/\sqrt3$ & 0 & & & & \cr
	& & $0$ & $1/\sqrt5$ & 0 & & & \cr
	& & & 0 & $1/\sqrt5$ & 0 & & \cr
	& & & & 0 & $1/\sqrt3$ & 0 &   \cr
	& & & & & $0$ & 1 & 0 \cr
\end{tabular}
\right)\,,
\eea
where the blank elements in the above matrices are zero. 

\end{document}